# The topological terms of interacting harmonic gravitational connections within the Maxwell-like gauge approach of gravity

**E Koorambas** [(1)*], **G Resconi** [(2)]

[(1)] Computational Applications, Group, Division of Applied Technologies, National Center for Science and Research 'Demokritos', Aghia Paraskevi-Athens, Greece

[(2)] Catholic University via Trieste, Theoretical Physics Dept. Brescia. Italy

(5 September 2017)

**Abstract.** In this paper, we base on the formalism of Symbolic Gauge Theory in the case of General relativity; we calculate the Feynman diagrams for the interaction between harmonic gravitational connections in the topological field theory. These calculations provide insights to interactions between gravity and dark energy. They could, therefore, have important repercussions for current cosmological problems.

## 1. Introduction

General relativity (GR), published by Albert Einstein in 1915 [1], generalizes special relativity and Newton's law of universal gravitation, providing a unified description of gravity as a geometric property of space and time, or space-time. The curvature of space-time, in particular, is directly related to the energy and momentum of whatever matter and radiation are present. This relation is specified by a system of partial differential equations, the Einstein field equations.

In the framework of geometric extensions of GR, a geometric unification between traditional gauge treatments of gravity, represented by a metric field, and dark energy, which arises as a corresponding gauge potential from the single SU (2) group has been proposed by the author [15]. Furthermore, the perturbation of gravitational waves caused by dark energy has been studied.

---

[*] elias.koor@gmail.com

Various workers have attempted to derive GR from a gauge-like principle, involving invariance of physics under transformations of the locally (i.e. in the tangent space at each point) acting Lorentz or Poincare group ([2], [3], [4]).

N. Wu (2003) [5−8] proposed a Quantum Gauge Theory of Gravity (QGTG) based on the gravitational gauge group (G). In Wu's theory, the gravitational interaction is considered as a fundamental interaction in a flat Minkowski space-time and not as space-time geometry.

A model of interacting massive gauge gravitons dark energy and a possible heavy gauge graviton dark matter resulting from shell decay of Higgs bosons have been developed recently by the author within the framework of QGTG [9, 10,11 12].

Symbolic Gauge Theory (SGT), a formalism applied to General Relativity (GR), was proposed by R. Mignani, E. Pessa and G. Resconi [13] and further developed by I.Licata and G. Resconi [14,15]. Manuel E. Rodrigues et al found black holes solutions within the framework of non-conservative theory of gravity [16]. The spherically symmetric solutions of this theory can be view as solutions that reproduce, the mass, the charge, the cosmological constant and the Rindler acceleration, without coupling with the matter content, i.e., in the vacuum. Application of SGT formalism in GR by the authors shows that the non-conservative gravitational equations can be represented by a wave equation with particular source where the variables are symmetric and anti-symmetric gravitational connections [17,18]. Cosmic acceleration can be derived by the harmonic gravitational connections (solutions of the wave equation). The proposed non-conservative theory of gravity can explain the observed variations of G at 5.9 year scale [19].

In this paper, we base on the formalism of SGT) [12-18], we calculate the Feynman diagrams for the interaction between harmonic gravitational connections in the topological field theory ([12],[20], [21], [22], [23], [24], [25], [26], [27], and [28]).

These calculations provide insights to interactions between gravity and dark energy. They could, therefore, have important repercussions for current cosmological problems.

## 2. The Maxwell-like gauge approach of gravity

Recently, G. Resconi, I. Licata and C. Gorda proposed a generalization of Gauge Theories based on the analysis of the structural characteristics of Maxwell theory, which can be considered as the 'prototype' of such theories (Maxwell-like) [14,18]. Theories of this type are based on a small number of principles related to different orders of commutators between covariant derivatives. We know that the Maxwell equations in the tensor form are

$$\partial_\mu F^{\mu\nu} = \frac{4\pi}{c} J^\nu$$
$$\partial^\alpha F^{\mu\nu} + \partial^\mu F^{\nu\alpha} + \partial^\nu F^{\alpha\mu} = 0, \tag{1}$$

where the controvariant four-vector which combines electric current density and electric charge density. $J^\nu = (cp, J_x, J_y, J_z)$ is the four-current, the electromagnetic tensor is $F^{\mu\nu} = \partial^\mu A_\nu - \partial^\nu A_\mu$. The four-potential $A^\mu = (\Phi, A_x, A_y, A_z)$ contains the electric potential and vector potential.

Using the covariant derivative notation defined by

$$D_\mu = \partial_\mu - ieA_\mu \tag{2}$$

we have the classic relation

$$[D_\mu, D_\nu] = D_\mu D_\nu - D_\nu D_\mu = -ieF_{\mu\nu}. \tag{3}$$

Thus, we have

$$F_{\mu\nu} = -F_{\nu\mu} \tag{4}$$

In order to write the extension of the Maxwell equations, we define the general commutators as in eq. (3) where $F_{\mu\nu}$ is the general form for any field generated by the transformation. Using eq. (3), the equation

$$[D_\mu, [D_\eta, D_\nu]]\psi + [D_\eta, [D_\nu, D_\mu]]\psi + [D_\nu, [D_\mu, D_\eta]]\psi = 0 \tag{5}$$

can be rewritten as

$$[D_\mu, F_{\eta\nu}]\psi + [D_\eta, F_{\nu\mu}]\psi + [D_\nu, F_{\mu\eta}]\psi = 0. \tag{6}$$

We also have the equation

$$[D_\gamma, [D_\alpha, D_\beta]]\psi = [D_\gamma, F_{\alpha\beta}]\psi = \chi J_{\gamma\alpha\beta}\psi \tag{7}$$

In conclusion, the Maxwell scheme for a general gauge transformation is

$$\begin{aligned}[D_\gamma, F_{\alpha\beta}] + [D_\alpha, F_{\beta\gamma}] + [D_\beta, F_{\gamma\alpha}] &= 0 \\ [D_\gamma, F_{\alpha\beta}] &= \chi J_{\gamma\alpha\beta}\psi,\end{aligned} \tag{8}$$

where $J_{\gamma\alpha\beta}$ are the currents of the particles that generate the gauge field F. Such currents have the conservation rule

$$J_{\underset{\gamma}{\alpha\beta}} + J_{\underset{\alpha}{\beta\gamma}} + J_{\underset{\beta}{\gamma\alpha}} = 0. \tag{9}$$

A modification of GR by Camenzind [29] was constructed as an analogy to the Yang-Mills theory, with the symmetry group SO (3.1). In the same vein, inspired by the foundations of gauge theory, Resconi et al. [12-14] constructed a modified theory of gravity in which the algebra of the covariant derivatives operators adds a term to the equations of motion. In the framework of this Maxwellian approach to gravity, we can obtain the general equation of motion by applying the commutator of $\nabla_\mu$ with the commutator of $\nabla_\alpha$ with $\nabla_\beta$ to a vector field $K_\nu$. This is achieved through the following steps:

First, it has been shown in [12] that

$$[\nabla_\mu, \nabla_\nu]V_\alpha = -R^\lambda_{\alpha\mu\nu}V_\lambda, \tag{10}$$

where the Riemann tensor is

$$R^\lambda_{\alpha\mu\nu} = \partial_\mu \Gamma^\lambda_{\nu\alpha} - \partial_\nu \Gamma^\lambda_{\mu\alpha} + \Gamma^\lambda_{\mu\alpha}\Gamma^\lambda_{\nu\sigma} - \Gamma^\lambda_{\nu\alpha}\Gamma^\lambda_{\mu\sigma}. \tag{11}$$

With application of the double commutator we have the dynamic equation

$$[\nabla_\mu,[\nabla_\alpha,\nabla_\beta]]K_\nu = (\nabla_\mu[\nabla_\alpha,\nabla_\beta])K_\nu - [\nabla_\alpha,\nabla_\beta](\nabla_\mu K_\nu)$$
$$= -(\nabla_\mu R^\lambda_{\nu\alpha\beta})K_\nu + R^\lambda_{\mu\alpha\beta}(\nabla_\lambda K_\nu), \tag{12}$$

where $\nabla_k$ is the covariant derivative and $K_\nu$ is the vacuum field. Using eq. (7), we connect the commutator with the gravity current in this way:

$$\chi J_{\mu\alpha\beta} K_\nu = (\nabla_\mu R^\lambda_{\nu\alpha\beta})K_\lambda + R^\lambda_{\mu\alpha\beta}(\nabla_\lambda K_\nu). \tag{13}$$

Then, for the conservation of the current we have after contractions the equation of motion

$$\nabla_\mu \left[ R^{\mu\nu} + k\left(T^{\mu\nu} + \frac{1}{2}g^{\mu\nu}T\right)\right]K_\nu + R^{\mu\nu}\left(\nabla_\mu K_\nu\right) = 0 \text{ [12]} \tag{14}$$

In the particular case where $\nabla_\mu K_\nu = 0$, considering the validity of Einstein's equations, we re-obtain the second Bianchi identity for a non-zero vector field $K_\nu$ [18]. These motion equations are stemming directly from a Lagrangian or variational principle, as in the case of Rastall theory [30]. All matter fields (scalar, vector and spinorial fields) are contained in the energy-momentum tensor, which gives us the interaction between matter and geometry (just as in RG).

Equation (16) can be interpreted as the equation of motion in a theory that can be viewed as a generalization of GR. This theory is derived from the foundations of gauge theory with an added term with respect to GR coming from the covariant derivatives commutator. We call this theory *Symbolic Gauge Theory of Gravity* (SGTG). The vector field $K_\nu$ can be interpreted as a field that describes the interaction with a substratum [12-13]. As in gauge theories, the vector field $K_\nu$ can be a gauge field of a certain local (as SU(2) gauge group for example) or global (as SU(3)) gauge group [18,16]. By contrast with gauge theories, however, in SGTG the vector field $K_\nu$ is a source of another operator, like an eigenvalue equation of the double commutator of covariant derivatives in [18]. Considering the validity of the Einstein Field Equations for the matter-sources, the double commutator acting on a non-vanishing vector field K can be expressed as a term proportional to a matter-source current J. The latter can in turn be related to the quadri-divergence of the Riemann tensor.

As is well known, through a contraction of the Bianchi identities, this quadri-divergence of the Riemann tensor can be expressed as a difference between quadri-divergences of the Ricci tensor. Using these expressions it is easy to see that *the quadri-divergence of the Riemann tensor vanishes for the Friedmann-Lemaitre-Robertson-Walker universe models* (FLRW): indeed, the space-time metric is highly symmetrical, and the only non-vanishing components of the Ricci tensor are those with equal indices.

This implies that the matter source term J vanishes for the FLRW models. Note that *the FLRW are conformally flat models* (the Weyl curvature tensor vanishes). In these cases there is indeed no non-uniformity of the gravitational field, in the sense that there are no tidal effects. In conclusion, in this context *the double commutator has to vanish for the FLRW models*.

Note, as well, that the de Sitter space-time solution also leads to a vanishing double commutator because of the high degree of symmetry of the model. On the other hand, the situation turns out to be very different for anisotropic and homogenous Bianchi-type models, as well as for inhomogeneous cosmological solutions. This occurs because the non-vanishing components of the Ricci tensor contain both diagonal and non-diagonal terms: in this way, the quadri-divergence of the Riemann tensor is non-zero and the double commutator is also non-vanishing. For Bianchi-type models, the Weyl curvature tensor is non-vanishing. As a consequence, these models are conformally non-flat. General inhomogeneous models can be shown to lead to a non-zero double commutator as well.

In addition, the Schwarzchild space-time solution also leads to a non-zero double commutator: these solutions are not conformally flat and thus are characterized by non-vanishing tidal effects.

The physical aspect which seems to be relevant for the interpretation of the double commutator is related to the *non-uniformity of the gravitational field*: if there are deep or fundamental requirements for a non-vanishing double commutator, this, in effect, makes the presence of non-vanishing tidal effects obligatory in physical models.

The dynamical equation (16) can be represented by a wave equation with particular source where the variables are symmetric and anti-symmetric gravitational connections (including torsion $T^\lambda_{\mu\nu} = \Gamma^\lambda_{\mu\nu} - \Gamma^\lambda_{\nu\mu}$ in one geometric picture). The dynamic equation for *non-conservative gravity* can be obtained by the SGT as follows [12-18]:

$$\left( \frac{\partial^2 \Gamma^\lambda_{\nu,\alpha}(x,t)}{\partial x^2} - \frac{\partial^2 \Gamma^\lambda_{\nu,\alpha}(x,t)}{\partial t^2} \right) V_\lambda = R_{\nu\alpha} - J_{\nu\alpha} \quad [12,13] \tag{15}$$

We thus obtain the interesting result that the derivative of the gravitational connection $\Gamma^\lambda_{\nu,\alpha}$ has wave behavior. The non-linear reaction of the self-coherent system produces a current that justifies the complexity of the gravitational field and non-linear properties of the non-metric gravitational waves. In analogy with non-linear optic, therefore, we can model gravitational phenomena as optic of tensor potentials. Equations (15) in the free field of the medium contain the Proca terms, $\Gamma_\rho \Gamma_\lambda$, the Chern-Simons terms ($\partial_\nu \Gamma_\rho$) $\Gamma_\lambda$, and the Maxwell-like terms ($\partial_\nu \Gamma_\rho$)($\partial_\mu \Gamma_\lambda$). In other words, we have the mass terms, the topologic terms and the electromagnetic field-like terms [31.32].

**3. The topological terms of interacting harmonic gravitational connections**

To investigate the interaction between the harmonic gravitational connections $\Gamma^a_\mu$ we use topological field theory. Our computation of the correlation function between the harmonic gravitational connections is analogous with that given in the literature ([28]).

$$\left\langle \int_{\gamma_2} dy^\nu \Gamma_{a\nu}^4 \int_{\gamma_1} dx^\mu \Gamma^b_{3\mu} \right\rangle_{S_{eff}} \tag{16}$$

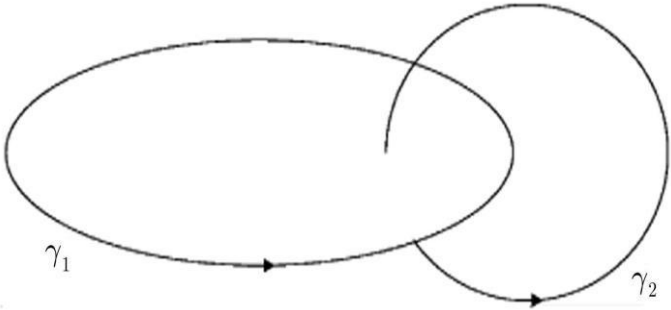

**Fig.1** Linking between $\gamma_1, \gamma_2$

For the case of the two harmonic gravitational connections $\Gamma^a_\mu$ and $\Gamma^a_\mu$ lying on two smooth, closed, non-intersecting curves $\gamma_1$ and $\gamma_2$ (Fig.1), the computation of the correlation function is reported up to 4-loop. This computation shows that the correlation function is unaffected by radiative correction. This result ensures the stability of the linking number with respect to the local perturbation ([28]). The local perturbation can be added to the Chern – Simons action given by

$$S_{eff} = \frac{g^{ac}}{2}\int d^3x \varepsilon^{\mu\nu\rho}\Gamma^a_{4\mu}\partial_\nu\Gamma^c_{3\rho} + \frac{\tau}{2!}\int d^3x R^{4\mu}_a R^a_{4\mu} R^{3\nu}_b R^b_{3\nu} \quad (17)$$

with $R^\mu_\alpha = \frac{1}{2}\varepsilon^{\mu\nu\rho} R_{\alpha\nu\rho}$ and $\tau$ being an arbitrary parameter with negative mass dimension, reflecting the power-counting non-renormalizability of the perturbation.

The Feynman diagrams for the interaction between harmonic gravitational connections in the topological field theory are similar to those described at [28]. To calculate the correlator function (16) we use the action

$$S_{eff} = \frac{1}{2}\int d^3x \varepsilon^{\mu\nu\rho}\Gamma^a_{4\mu}\partial_\nu\Gamma^c_{3\rho} + \int d^3x b_c \partial^\rho \Gamma^c_{3\rho} + \frac{\tau}{4!}\int d^3x : R^{4\mu}_a R^a_{4\mu} R^{3\nu}_b R^b_{3\nu} : ,...... \quad (18)$$

The Lagrange multiple $b$ is introduced to this function to implement the Landau gauge.

To evaluate the Feynman diagrams we use the following equations:

$$\partial^2 \frac{1}{|x-y|} = -4\pi\delta^3(x-y),$$
$$\left\langle \Gamma^a_{4\mu}(x), \Gamma^b_{3\nu}(y) \right\rangle_{S_{eff}} = \frac{g^{ab}}{4\pi}\varepsilon_{\mu\nu\rho}\frac{(x-y)^\rho}{|x-y|^3} \quad (19)$$

From equations (19) we obtain

$$\left\langle \Gamma^a_{4\mu}(x), R^b_{3\nu}(y) \right\rangle = g^{ab} g_{\mu\nu}\delta(x-y) + \partial_\mu\partial_\nu \frac{1}{4\pi|x-y|},$$
$$\left\langle \bar{R}^a_{4\mu}(x), R^b_{3\nu}(y) \right\rangle = -\partial^{ab}_{\mu\nu}\delta^3(x-y) \quad (20)$$

where

$$\partial^{ab}_{\mu\nu} = g^{ab}\varepsilon_{\mu\nu\rho}\partial^\rho \quad (21)$$

the transverse derivative operator.

The Feynman diagrams that contribute to the correlation function (16) are of two-loop order (see Figure 2). Therefore, they correspond to the following integral:

$$I^{(2)} = \int_{\gamma_1} dx^\mu \int_{\gamma_2} dy^\nu \int d^3z_1 d^3z_2 \left\langle \Gamma^a_\mu(x) R^b_\alpha(z_1) \right\rangle \left\langle \Gamma_{\nu a}(y) R_{\gamma b}(z_2) \right\rangle \left\langle R^a_\beta(z_1) R^b_\rho(z_2) \right\rangle$$
$$[4\left\langle \bar{R}^\alpha_a(z_1) \bar{R}^\beta_b(z_2) \right\rangle \left\langle \bar{R}^{a\beta}(z_1) \bar{R}^{b\gamma}(z_2) \right\rangle + 2\left\langle \bar{R}^\alpha_a(z_1) \bar{R}^\gamma_b(z_2) \right\rangle \left\langle \bar{R}^{a\beta}(z_1) \bar{R}^{b\rho}(z_2) \right\rangle] \quad (22)$$

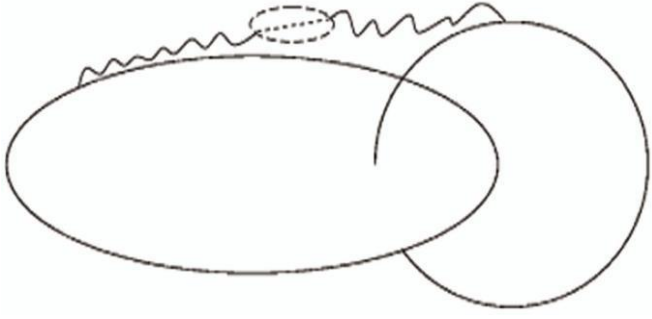

**Fig.2** Two loop contribution.

Let us analyze the first term of the above expression. Making use of the propagators (20), we obtain

$$-4\int_{\gamma_1} dx^\mu \int_{\gamma_2} dy^\nu \int d^3z_1 d^3z_2 [g^{ab} g_{\mu\alpha}\delta^3(x-z_1) + \partial_\mu \partial_a \frac{1}{4\pi|x-z_1|}]$$
$$[g_{ab} g_{\nu\gamma}\delta^3(y-z_2) + \partial_\nu \partial_\gamma \frac{1}{4\pi|y-z_2|}] \qquad (23)$$
$$[\partial^{ab}_{\beta\rho}\delta^3(z_1-z_2)][\partial^{\alpha\rho}_{ab}\delta^3(z_1-z_2)][\partial^{\beta\gamma ab}\delta^3(z_1-z_2)]$$

The terms containing the derivatives $\partial_\mu$ and $\partial_\nu$ do not contribute, as they correspond to total derivatives on closed curves. The term described by expression (23) then becomes

$$4\int_{\gamma_1} dx^\mu \int_{\gamma_2} dy^\nu \int d^3z_1 d^3z_2 \delta^3(x-z_1)\delta^3(y-z_2)[\partial^{ab}_{\beta\rho}\delta^3(z_1-z_2)]$$
$$[\partial^{\rho a}_\mu \delta^3(z_1-z_2)][\partial^\beta_{a\ \nu}\delta^3(z_1-z_2)] \qquad (24)$$

This can be also obtained by regularizing the delta functions with coinciding arguments through the point-splitting procedure already used by Polyakov [33]:

$$\delta_\varepsilon(z_1-z_2) = \frac{1}{(2\pi\varepsilon)^{3/2}} e^{-(z_1-z_2)^2/2\varepsilon} \qquad (25)$$

More precisely, whenever a product of n delta functions with coinciding arguments occurs, it is understood as

$$[\delta^3(z_1-z_2)]^n = [\delta_\varepsilon(z_1-z_2)]^{n-1}\delta^3(z_1-z_2) \qquad (26)$$

where the limit $\varepsilon \to 0$ is meant to be taken at the end of all calculations. Expression (24) becomes,

$$I^{(2)} = 4\lim_{\varepsilon\to 0}\int_{\gamma_1} dx^\mu \int_{\gamma_2} dy^\nu \int d^3z_1 d^3z_2 \delta^3(x-z_1)\delta^3(y-z_2)[\partial^{ab}_{\beta\rho}\delta_\varepsilon(z_1-z_2)]$$
$$[\partial^{\rho a}_\mu \delta_\varepsilon(z_1-z_2)][\partial^\beta_{a\ \nu}\delta^3(z_1-z_2)] \qquad (27)$$

Whatever the order of integration, we obtain, an expression containing $\delta^3(x-y)$, which leads to a null result.

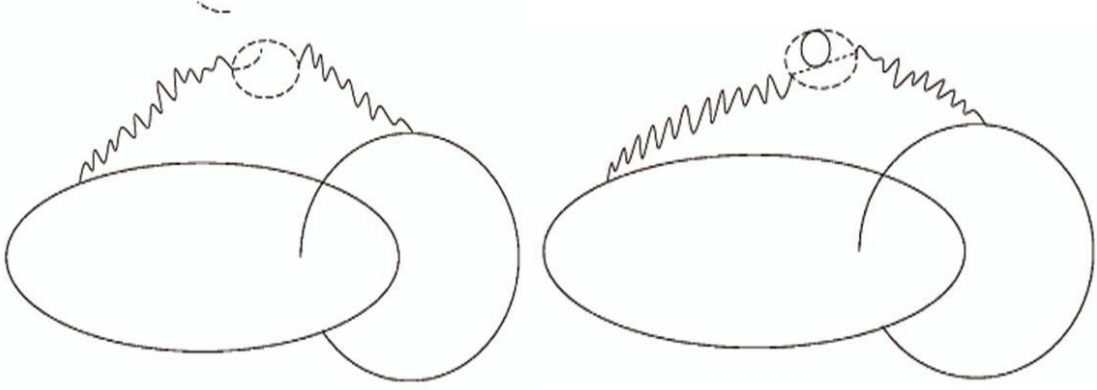

**Fig.3** Three loop contribution. **Fig. 4** Four loop contribution.

Analysis of the second term of (22) gives similar results. The two-loop diagram of Fig.2, therefore, does not contribute to the correlator (16). Concerning the higher-order contributions in the perturbation theory, the results are of a similar nature. The topologically distinct diagrams contributing to the 3- and 4-loop are given in Figs. 3, 4 and 5. For instance, a typical contraction from Fig.3 is proportional to

$$I^{(3)} = \int_{\gamma_1} dx^\mu \int_{\gamma_2} dy^\nu \int d^3z_1 d^3z_2 d^3z_3 g^{ab} g_{\mu\alpha} \delta^3(x-z_1) g^{ab} g_{\nu\gamma} \delta^3(y-z_2)$$

$$[\partial^{\alpha\gamma}_{ab}\delta^3(z_1-z_2)][\partial^{ab}_{\beta\delta}\delta^3(z_1-z_3)][\partial^{\beta\lambda}_{ab}\delta^3(z_1-z_3)][\partial^{ab}_{\rho\lambda}\delta^3(z_2-z_3)][\partial^{\rho\delta}_{ab}\delta^3(z_2-z_3)]$$

(28)

$$= \int_{\gamma_1} dx^\mu \int_{\gamma_2} dy^\nu \int d^3z_3 [\partial_{\mu\nu\, ab}\delta^3(x-y)][\partial^{ab}_{\beta\delta}\delta_\varepsilon(x-z_3)][\partial^{\beta\lambda}_{ab}\delta^3(x-z_3)]$$

$$[\partial^{ab}_{\rho\lambda}\delta_\varepsilon(y-z_3)][\partial^{\rho\delta}_{ab}\delta^3(y-z_3)]$$

while the diagram of Fig.4 gives

$$I^{(4)} = \int_{\gamma_1} dx^\mu \int_{\gamma_2} dy^\nu \int d^3z_1....d^3z_4 g^{ab} g_{\mu\alpha} \delta^3(x-z_1) g_{ab} g_{\nu\gamma} \delta^3(y-z_2)$$

$$[\partial^{\alpha\gamma}_{ab}\delta^3(z_1-z_2)][\partial^{\beta\lambda}_{ab}\delta^3(z_1-z_3)][\partial^{ab}_{\beta\sigma}\delta^3(z_1-z_4)]$$

$$[\partial^{\rho\sigma}_{ab}\delta^3(z_2-z_4)][\partial^{ab}_{\rho\lambda}\delta^3(z_2-z_3)][\partial^{\varphi\omega}_{ab}\delta^3(z_3-z_4)][\partial^{ab}_{\varphi\omega}\delta^3(z_3-z_4)]$$

(29)

$$= \int_{\gamma_1} dx^\mu \int_{\gamma_2} dy^\nu \int d^3z_3 d^3z_4 [\partial_{\mu\nu\, ab}\delta^3(x-y)][\partial^{\beta\lambda}_{ab}\delta_\varepsilon(x-z_3)][\partial^{ab}_{\beta\sigma}\delta^3(x-z_4)]$$

$$[\partial^{\rho\sigma}_{ab}\delta^3(y-z_3)][\partial^{ab}_{\rho\lambda}\delta^3(y-z_3)][\partial^{\varphi\omega}_{ab}\delta_\varepsilon(z_3-z_4)][\partial^{ab}_{\varphi\omega}\delta^3(z_3-z_4)]$$

All terms in all possible diagrams may are, then, shown to be proportional to $\delta^3(x-y)$ (or its derivatives). One may easily convince oneself that this mechanism also applies to any order in perturbation theory. As it is always $x \neq y$, all these diagrams amount to a null correction to the basic diagram, so that the correlation function (16) for two closed smooth nonintersecting curves $\gamma_1$ and $\gamma_2$ gives their linking number to all orders:

$$\left\langle \Gamma^b_{3\mu}(x), \Gamma^a_{4\nu}(y) \right\rangle_{S_{eff}} = \chi(\gamma_1, \gamma_2) \qquad (30)$$

## 4. Conclusion

Equation (24) links harmonic gravitational connections, $\Gamma^a_\beta$. This expression of the interaction between harmonic gravitational connections may have potentially very important implications to cosmology. We find that the interactions between harmonic gravitational connections are independent of metric. These interactions are, therefore, within the framework of perturbing topological field theory. The topological properties of these interactions are represented by knots and links. The size, exact shape, location etc of these knots and links are not of immediate concern for the problem at hand (interaction between dark mater/energy and gravitons). Equation (30) depends only on the topological relationship of the knots with each other. This invariant may have a physical interpretation: it may represent the work done to move a graviton (dark energy) around one knot in three dimensional space while a graviton runs around the other knot.